\documentclass[aps,prl,twocolumn,10pt,superscriptaddress,nofootinbib,floatfix]{revtex4-1}

\usepackage{amsmath}
\usepackage{bm}
\usepackage{amsfonts}
\usepackage{graphicx}
\usepackage{epstopdf}
\usepackage{hyperref}
\usepackage{array}
\usepackage{soul}
\usepackage{color}
\usepackage{booktabs}
\hypersetup{colorlinks   = true,
            urlcolor     = blue,
            citecolor    = blue,
            linkcolor    = blue,
            menucolor    = blue,
            anchorcolor  = blue,
            filecolor    = blue}

\begin{document}

\title{Atomic Quantum Sensors for High-Frequency Gravitational Wave Searches}

\author{Yi-Fu Cai}
\email{yifucai@ustc.edu.cn}
\affiliation{Department of Astronomy, School of Physical Sciences,
University of Science and Technology of China, Hefei, Anhui 230026, China}
\affiliation{CAS Key Laboratory for Researches in Galaxies and Cosmology,
School of Astronomy and Space Science, University of Science and Technology of China, Hefei, Anhui 230026, China}

\author{Luca Visinelli}
\email{lvisinelli@unisa.it}
\affiliation{Dipartimento di Fisica ``E.R.\ Caianiello'', Universit\`a degli Studi di Salerno,\\ Via Giovanni Paolo II, 132 - 84084 Fisciano (SA), Italy}
\affiliation{Istituto Nazionale di Fisica Nucleare - Gruppo Collegato di Salerno - Sezione di Napoli,\\ Via Giovanni Paolo II, 132 - 84084 Fisciano (SA), Italy}

\author{Sheng-Feng Yan}
\email{sfyan@uestc.edu.cn \newline }
\affiliation{School of Physics, The University of Electronic Science and Technology of China, 88 Tian-run Road, Chengdu, China}
\affiliation{College of Physics, Chengdu University of Technology, Chengdu 610059, China}

\begin{abstract}
High-frequency gravitational waves (GWs), spanning frequencies from the microwave to the optical band, remain experimentally unexplored despite strong motivation from early-Universe dynamics, high-energy cosmology, and exotic compact objects. We propose a detection framework in which an incident GW excites an eigenmode of a high-$Q$ resonator in the presence of a static magnetic field through GW-induced electromagnetic mode conversion; the resulting cavity field is then read out using atomic sensors placed outside the magnetized volume. We analyze two concrete architectures: microwave detection based on Rydberg transitions and optical/near-infrared Raman schemes. For each, we derive projected strain sensitivities achievable with realistic, though ambitious, magnetic fields, cavity parameters, and atomic ensembles. Under optimistic assumptions on cavity performance, signal coherence, and technical noise, optical Raman implementations could approach benchmark narrowband coherent strain sensitivities relevant for speculative high frequency GW scenarios, while microwave systems may probe benchmark sensitivities in an otherwise unexplored frequency range. These setups motivate advances in high-$Q$ cavities, strong-field magnets, and quantum-limited atomic sensors, with broader implications for quantum instrumentation and fundamental physics.
\end{abstract}

\maketitle

\textbf{\textit{Introduction ---}} The direct detection of gravitational waves (GWs) has opened a new observational window on the Universe and established GW astronomy as a precision tool for astrophysics and cosmology~\cite{LIGOScientific:2016aoc}. Present observatories, however, probe only a limited frequency range: ground-based interferometers access the audio band from tens to thousands of Hz~\cite{LIGOScientific:2007fwp, Accadia:2010zzc, Punturo:2010zz, Hall:2022dik, KAGRA:2022twx}, planned space missions target the mHz regime~\cite{Jennrich:2021okh, Baker:2019pnp, Lu:2019log, Luo:2019zal, Bailes:2021tot, Cai:2023ywp, Ren:2023yec, Wang:2025tvw, Wang:2024tnk}, and pulsar timing arrays access the nHz domain~\cite{Kerr:2020qdo, NANOGrav:2023gor, EPTA:2023fyk, Kelley:2025yud}. Atom interferometer proposals aim to bridge the intermediate sub-Hz to 10\,Hz band using quantum superpositions of atomic states~\cite{Graham:2012sy, Graham:2017pmn, Badurina:2019hst, AEDGE:2019nxb, Bauer:2024hfv}. Together, these efforts have revolutionized our understanding of black holes (BHs), neutron stars, and multi-messenger astronomy, providing a powerful picture of the dynamic Universe~\cite{Bian:2025ifp, Sathyaprakash:2009xs, Caprini:2018mtu}.

By contrast, the high-frequency gravitational waves (HFGWs) domain, spanning frequencies above the MHz scale, remains essentially unexplored~\cite{Aggarwal:2020olq, Aggarwal:2025noe}. Conventional astrophysical processes are not expected to produce detectable signals in this regime~\cite{Gatti:2024mde, Amaral:2026bef}. Nonetheless, a variety of speculative early-Universe mechanisms, including violent phase transitions~\cite{Jinno:2016knw, Okada:2018xdh, Huang:2020bbe, Okada:2020vvb, Nakai:2020oit, Addazi:2023jvg, Kushwaha:2022twx}, decays of topological defects~\cite{Damour:2001bk, Leblond:2009fq, Jones-Smith:2007hib}, exotic compact object dynamics~\cite{Abdikamalov:2020jzn, Blas:2022xco, Cardoso:2019rvt}, or binary mergers of light primordial BHs~\cite{Bavera:2021wmw, Carr:2023tpt, LISACosmologyWorkingGroup:2023njw, Dong:2015yjs, Yin:2026hvw}, could in principle contribute to the background at these frequencies. Detecting HFGWs would test uniquely high-energy scales and epoch inaccessible by other means. Even null results yield non-trivial bounds on physics beyond the Standard Model.

Detecting HFGWs is exceptionally challenging. Standard Michelson interferometry fails at short wavelengths because of arm-length mismatch, motivating alternative approaches. One long-studied mechanism is the Gertsenshtein effect~\cite{Gertsenshtein1962}, where a GW traversing a static magnetic field converts into an electromagnetic (EM) wave. Although the conversion probability is tiny, it scales with the square of both the field strength and interaction length, suggesting that strong fields and resonant structures could amplify the signal~\cite{Chen:1994ch, Harry:1996gh, Cillis:1996qy, Li:2003tv, Aguiar:2010kn, Berlin:2021txa, Liu:2023mll, Ito:2023fcr, Ito:2023nkq, Capdevilla:2024cby, Capdevilla:2025omb, Kushwaha:2025mia, Schenk:2025ria}.

In this {\it Letter}, we introduce a cascade approach in which GW--EM conversion in a resonant cavity~\cite{Gertsenshtein1962} is interfaced with atomic quantum sensors~\cite{Ye:2008rva}, providing a new avenue to probe the HFGW band. A passing HFGW interacting with a strong static magnetic field generates a cavity-enhanced EM wave; this signal imprints sidebands, frequency shifts, or phase modulations on a probe laser for Raman schemes, with the high-$Q$ cavity acting simultaneously as transducer and amplifier. Drawing inspiration from cavity optomechanics~\cite{Aspelmeyer:2013lha}, where mechanical vibrations modulate optical fields, this scheme highlights the potential of high-$Q$ resonators and quantum sensors~\cite{Kasevich:1992yii} to bridge general relativity, quantum optics, and precision metrology. The cavity field couples to atoms via the electric dipole interaction, enabling long-lived coherences, large ensembles for projection-noise suppression, and coherent phase ~\cite{Fischer:1994ed, Kanno:2023whr}. Raman two-photon schemes in atoms such as cesium (Cs), Rb, Sr, and Yb offer a natural interface by coherently coupling electronic ground states while suppressing spontaneous emission. Complementary strategies using Rydberg states in the microwave band, photoionization thresholds in the near-infrared (NIR), and narrow inner-shell resonances extend sensitivity across a broad frequency range, possibly converting ultraweak GW signals into measurable quantum observables.

We develop a framework for cavity–atomic HFGW searches, with the conservative benchmark representing the most experimentally grounded scenario. Rather than starting from the vacuum graviton--photon conversion probability, we treat the GW as an effective source for a dissipative cavity eigenmode in the presence of a static magnetic field, and then analyze how the resulting cavity field couples to atomic quantum sensors. The cavity response is mapped to transition and Rabi rates, and shot-noise-limited sensitivities are estimated for representative microwave and optical readout architectures. Our goal is to clarify where atomic quantum sensors might realistically contribute to HFGW searches, and which future advances in cavity technology, magnetic transduction, or atomic control could most improve prospects in this unexplored frequency band. We adopt natural units $c=\hbar=1$ unless otherwise noted, while $G$ is Newton's constant.

\begin{figure}[htb!]
    \centering
    \includegraphics[width=\linewidth]{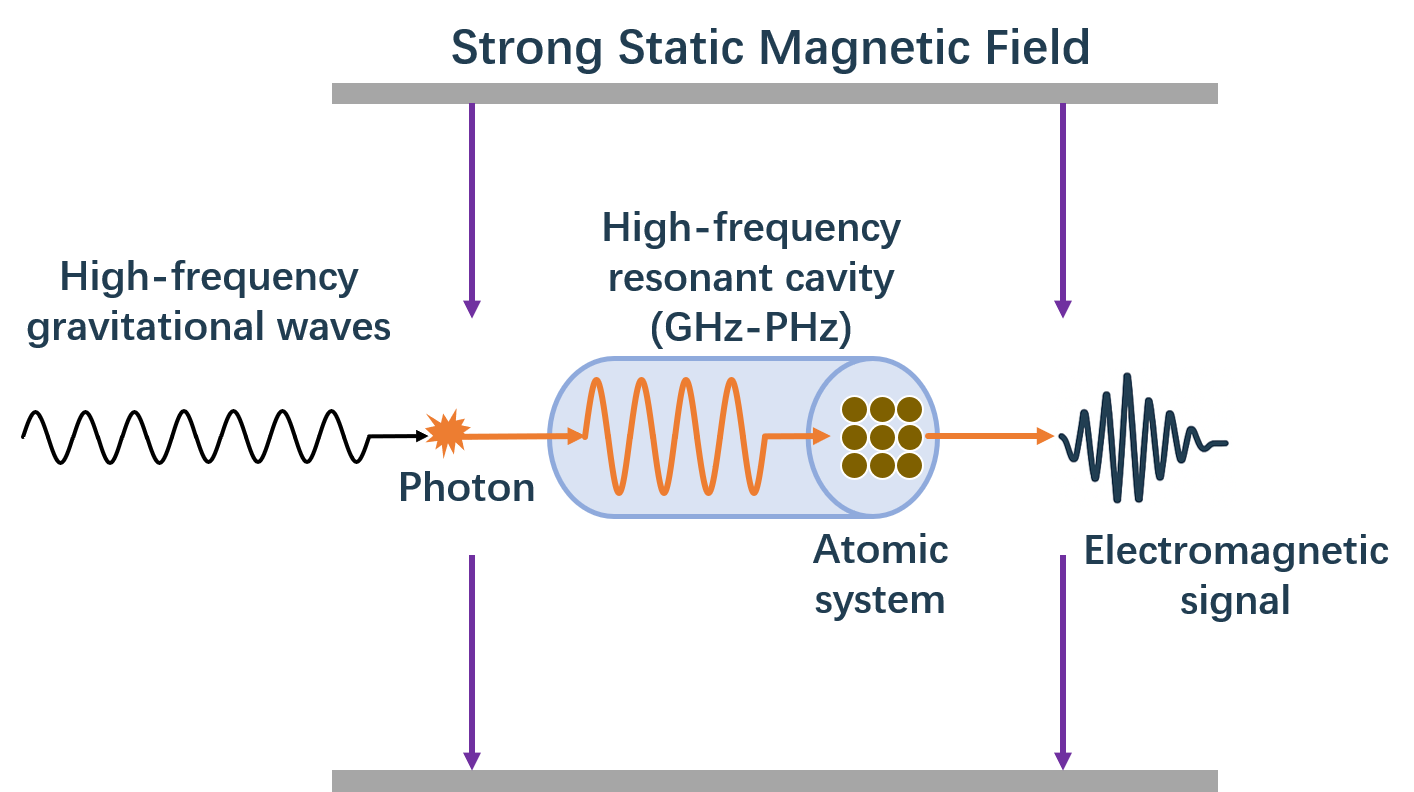}
    \caption{Schematic of the detection concept. A HFGW traversing a static magnetic field excites a resonant EM mode of a high-$Q$ cavity via GW-induced EM conversion, with the mode volume overlapping the magnetized region. The cavity field is then coupled to atomic quantum sensors located outside the high-field region: either directly through microwave transitions in Rydberg atoms, or indirectly via Raman processes involving an auxiliary laser. The resulting atomic response provides a measure of the GW strain spectral density.}
    \label{fig:Sketch}
\end{figure}

\textbf{\textit{Methods ---}} We describe the detector as a resonant EM cavity driven by a GW in the presence of a static magnetic field, followed by quantum-limited readout using atomic sensors, see Fig.~\ref{fig:Sketch}. This cascade formulation treats the cavity as an active transducer, explicitly incorporating its mode structure, dissipation, and finite coherence time, rather than relying on graviton-photon conversion in vacuum.

Starting from the Maxwell--Einstein action,
\begin{equation}
    S = \int {\rm d}^4x \sqrt{-g}\left(-\frac14 g_{\mu\alpha}g_{\nu\beta}F^{\mu\nu}F^{\alpha\beta}\right),
\end{equation}
and linearizing around flat spacetime, $g_{\mu\nu}=\eta_{\mu\nu}+h_{\mu\nu}$ with $|h_{\mu\nu}|\ll 1$, the GW perturbation induces an effective current for the EM field in the presence of a static magnetic field $B$. Expanding the EM field in cavity eigenmodes, the amplitude of the $n$-th mode takes the driven-oscillator form~\cite{Berlin:2021txa, Berlin:2023grv, Navarro:2023eii}
\begin{equation}
    e_n(\omega)
    = -\,i\,\omega\,B\,V^{5/6}\,\mathcal{T}_n(\omega)\,
    \sum_A h_A(\omega)\,\eta_A \, ,
    \label{eq:en_methods}
\end{equation}
where $V$ is an effective volume scale set by the mode normalization, $h_A(\omega)$ are the GW polarization amplitudes, $\eta_A$ are dimensionless overlap coefficients encoding the projection of the GW-induced current onto the cavity mode, and $\mathcal{T}_n(\omega)$ is the cavity transfer function. This formulation consistently incorporates boundary conditions, mode structure, and dissipation. Similar treatments also emphasize the role of coherence and geometry~\cite{Ringwald:2020ist}. Near resonance, $\omega\simeq\omega_n=2\pi f$, the response is well approximated by
\begin{equation}
    \mathcal{T}_n(\omega)
    \simeq
    \left[
    \left(1-\frac{\omega_n^2}{\omega^2}\right)
    - i\,\frac{1}{Q}\frac{\omega_n}{\omega}
    \right]^{-1},
    \label{eq:T_methods}
\end{equation}
so that both resonant enhancement and dissipation are treated consistently within the cavity dynamics. For a coherent narrowband source centered on resonance, this leads to the signal power estimate
\begin{equation}
    P_{\rm sig}
    \simeq
    \frac14\,B^2\,V^{5/3}\,\eta^2\,\omega_n^3\,h_0^2\,Q_{\rm eff},
    \label{eq:Psig_methods}
\end{equation}
where $\eta^2=\sum_A \eta_A^2$ and
\begin{equation}
    Q_{\rm eff} \equiv \min(Q, N_{\rm cyc})
\end{equation}
accounts for the finite coherence of the GW signal over the cavity response time~\cite{Gatti:2024mde}. Resonant enhancement requires a sufficiently coherent, narrowband signal: for rapidly chirping or stochastic sources the effective gain is correspondingly reduced.

Expressing the signal in terms of the strain spectral density, $h_0^2 \simeq S_h(f)\Delta f$, with bandwidth
\begin{equation}
    \label{eq:Deltaf_methods}
    \Delta f = \min\!\left(\frac{1}{\tau}, \frac{f}{2Q}\right),
\end{equation}
we obtain
\begin{equation}
    P_{\rm sig}
    \simeq
    \frac14\,(2\pi)^3\,B^2\,V^{5/3}\,\eta^2\,f^3\,
    Q_{\rm eff}\,S_h(f)\,\Delta f .
    \label{eq:Psig_Sh_methods}
\end{equation}
All sensitivities assume optimal alignment between the GW propagation direction, polarization, and cavity mode. Sky and polarization averages weaken sensitivities, potentially by more than an order of magnitude. The resulting cavity field provides the input to the atomic readout. The effective electric field amplitude in the cavity mode is
\begin{equation}
    E_{\rm cav}^2
    \simeq
    \frac{2\,P_{\rm sig}\,Q}{\omega_n\,V_{\rm mode}} \, ,
\end{equation}
where $V_{\rm mode}$ is the mode volume. Substituting Eq.~\eqref{eq:Psig_Sh_methods} yields
\begin{equation}
    E_{\rm cav}^2
    \simeq
    (2\pi)^2\,B^2\,V^{2/3}\,\mathcal{G}
    \, f^2\,Q\,Q_{\rm eff}\,S_h(f)\,\Delta f \,,
    \label{eq:Ecav_methods}
\end{equation}
where $\mathcal{G} \equiv \eta^2\,V/V_{\rm mode}$. Setting $\eta\sim 0.14$ and $V_{\rm mode}\sim V$ gives $\mathcal{G} \sim 0.02$, which we take as the benchmark value. Since the strain sensitivity scales as $\sqrt{S_h}\propto \mathcal{G}^{-1/2}$, slight variations in the geometric overlap factor do not sensibly affect the projected reaches.

The coupling to matter proceeds in two complementary ways. For microwave detection using Rydberg atoms or superconducting qubits, the cavity field directly drives dipole transitions. The single-particle Rabi frequency is $\Omega_{\rm sp} = d_{\rm Ryd} E_{\rm cav}$, where $d_{\rm Ryd}$ is the transition dipole moment. For an ensemble of $N$ atoms interrogated over a time $\tau$, the shot-noise-limited strain sensitivity follows from $\Omega_{\rm sp}\tau\sqrt{N}\sim 1$, giving
\begin{equation}
    \sqrt{S_{h,{\rm min}}^{\rm (Rydberg)}}
    \simeq
    \frac{1}{2\pi B d_{\rm Ryd}\tau}
    \left[\mathcal{G}\,V^{2/3}\,Q\,Q_{\rm eff}\,N\,f^2\,\Delta f\right]^{\!-\frac{1}{2}} \, .
    \label{eq:Sh_Ryd_methods}
\end{equation}
In optical/NIR Raman schemes, the cavity field provides one or both legs of a two-photon transition between long-lived ground states. In a mixed configuration with a classical control laser of field amplitude $E_{\rm laser}$, the effective Rabi frequency is
\begin{equation}
    \Omega_{2\gamma}^{(I)}
    = \frac{d_{\rm opt}^2}{2\Delta}\,E_{\rm laser}\,E_{\rm cav},
\end{equation}
leading to the sensitivity
\begin{equation}
    \sqrt{S_{h,{\rm min}}^{({\rm Raman}, I)}}
    \simeq
    \frac{\Delta}{\pi B d_{\rm opt}^2 E_{\rm laser}\tau}
    \left[\mathcal{G}\,V^{2/3}\,Q\,Q_{\rm eff}\,Nf^2\Delta f\right]^{\!-\frac{1}{2}} \, .
    \label{eq:Sh_Raman_methods}
\end{equation}
A configuration in which both Raman photons originate from the cavity field yields a weaker scaling, $\Omega_{2\gamma}\propto E_{\rm cav}^2$, and correspondingly reduced sensitivity.

In all cases, the atomic sensor probes the EM excitation generated in the cavity by the GW, while the overall sensitivity is controlled by the efficiency of the GW-to-cavity transduction and by the coherence properties of the signal through $Q_{\rm eff}$. The interrogation time $\tau$ is limited by decoherence, including cavity losses, thermal occupation, spontaneous scattering in Raman schemes, and technical noise.

\begin{figure}[htb!]
    \centering
    \includegraphics[width=\linewidth]{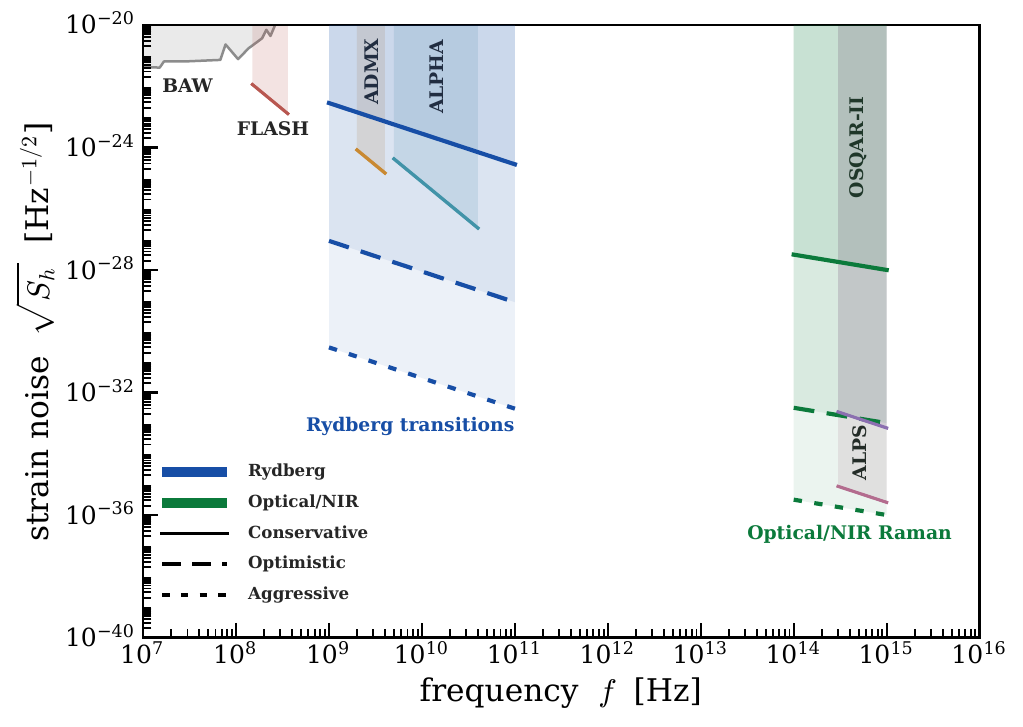}
    \caption{Benchmark shot-noise-limited narrowband strain sensitivities $\sqrt{S_{h,\rm min}}$ for coherent GW-induced cavity excitation with atomic readout. Solid curves show microwave Rydberg (blue) and optical/NIR Raman schemes (green); shaded bands span conservative, optimistic, and aggressive assumptions. Also shown for comparison are existing and projected sensitivities from BAW~\cite{Goryachev:2014yra, Goryachev:2021zzn, Campbell:2023qbf, Campbell:2025mks}, FLASH~\cite{Alesini:2019nzq, Alesini:2023qed}, ALPHA~\cite{Lawson:2019brd, ALPHA:2022rxj}, ALPS-II~\cite{Bahre:2013ywa, Beacham:2019nyx}, ADMX~\cite{ADMX:2018gho, ADMX:2019uok, ADMX:2021nhd, Chakrabarty:2023rha}, and OSQAR-II~\cite{OSQAR:2007oyv, Ballou:2014myz}.}
    \label{fig:reach}
\end{figure}

\textbf{\textit{Results ---}} In this work, the projected reaches for the atomic readout architectures are interpreted as benchmark sensitivities to narrowband coherent signals within the cavity bandwidth. They are not matched to a specific established source model for persistent coherent HFGWs, and comparisons with stochastic-background or broadband limits should be understood accordingly. Under comparable narrowband coherent-signal assumptions, the projected sensitivities for microwave and optical/NIR schemes can exceed those of current resonant mass detectors and EM oscillators, highlighting the potential of atomic sensors for HFGW searches. We examine the feasibility of microwave and optical/NIR implementations in three representative regimes: a conservative baseline reflecting current capabilities, an optimistic extrapolation based on plausible near-future advances, and an aggressive scenario exploring idealized limits with spin squeezing and extended coherence times. The corresponding parameter choices are summarized in Table~\ref{table:configurations}. The sensitivity projections in Table~\ref{table:configurations} and Fig.~\ref{fig:reach} assume unit conversion and readout efficiency, and optimal alignment between the GW propagation direction, the static magnetic field, and the cavity mode.

\begin{table}[htb!]
\centering
\begin{tabular}{c|c|c|c}
\hline
\toprule
\textbf{Quantity/Setup} & \textbf{Con.} & \textbf{Opt.} & \textbf{Agg.} \\
\hline \hline
Magnetic field $B$ & $1\,{\rm T}$ & $10\,{\rm T}$ & $30\,{\rm T}$ \\
Effective cavity volume $V$ & $10^{-2}\,{\rm m}^3$ & $10^{-1}\,{\rm m}^3$ & $1\,{\rm m}^3$ \\
Quality factor $Q$ & $10^6$ & $10^9$ & $10^{10}$ \\
Number of atoms $N$ & $10^6$ & $10^8$ & $10^9$ \\
Interrogation time $\tau$ & $10\,\mu{\rm s}$ & $1\,{\rm ms}$ & $10\,{\rm ms}$ \\
Squeezing & No & No & $20\,{\rm dB}$ \\
\bottomrule
\hline
\end{tabular}
\caption{Benchmark configurations used in the sensitivity projections: conservative (Con.), optimistic (Opt.), and aggressive (Agg.). We set $\mathcal G = 0.02$ as a representative geometric overlap factor.}
\label{table:configurations}
\end{table}

For microwave transitions in Rydberg atoms or superconducting qubits, the relevant sensitivity is given by Eq.~\eqref{eq:Sh_Ryd_methods}. Transitions in the $(1$--$100)$\,GHz band are naturally accessible and compatible with high-$Q$ superconducting cavities, enabling long coherence times, large dipole moments, and strong magnetic fields. Detection proceeds via coherent Rabi excitation, read out through fluorescence or cavity response, while a static magnetic field mediates graviton--photon conversion via the Gertsenshtein effect. As a representative case, consider a baseline prototype operating at $f=10\,$GHz with $N$ Rydberg atoms of dipole $d_{\rm Ryd}\simeq 10^{-26}\,\mathrm{C\,m}$. The conservative configuration yields a SSD $\sqrt{S_{h, {\rm min}}} \sim 5\times10^{-25}{\rm\,Hz^{-\frac{1}{2}}}$. In the optimistic configuration, this improves to $\sqrt{S_{h, {\rm min}}} \sim 10^{-29}{\rm\,Hz^{-\frac{1}{2}}}$, and in the aggressive limit with squeezing and long interrogation times, the sensitivity could reach $\sqrt{S_{h, {\rm min}}} \sim 10^{-32}{\rm\,Hz^{-\frac{1}{2}}}$. These sensitivities apply to narrowband, coherent GW signals resonantly enhanced within the cavity bandwidth $\Delta f$, assuming $Q_{\rm eff}\sim Q$.

As an alternative to Rydberg implementations, one may consider laser-driven optical/NIR Raman transitions, where two long-lived ground states are coupled via a virtually populated excited state, giving access to GW frequencies in the optical band. Raman schemes benefit from large atomic ensembles with well-characterized transitions. For concreteness, we focus on the Cs D2 line~\cite{Cs}, with natural linewidth $\Gamma/2\pi \approx 5.2\,$MHz for the 852\,nm transition and effective dipole $d_{\rm opt} \simeq 2.6\times 10^{-29}\,\mathrm{C\,m}$. With a laser field amplitude $E_{\rm laser} = 10^5$\,V/m and a single-photon detuning $\Delta = 2\pi$\,GHz, Eq.~\eqref{eq:Sh_Raman_methods} yields $\sqrt{S_{h, {\rm min}}} \sim 10^{-28}{\rm\,Hz^{-\frac{1}{2}}}$ for $f = 10^{15}$\,Hz in a conservative configuration, improving the sensitivity to $\sqrt{S_{h, {\rm min}}} \sim 10^{-33}{\rm\,Hz^{-\frac{1}{2}}}$ in the optimistic regime. In the aggressive configuration, the sensitivity could reach $\sqrt{S_{h, {\rm min}}} \sim 10^{-36}{\rm\,Hz^{-\frac{1}{2}}}$. GW backgrounds contribute incoherently through their strain power spectral density within the detector bandwidth fixed by the cavity quality factor~\cite{Romano:2016dpx}. These benchmark coherent sensitivities can approach cosmological benchmark curves under optimistic assumptions on coherence, cavity performance, and noise. Achieving such performance, however, requires stringent control of technical noise and decoherence. Quantum limitations demand that noise fluctuations of the effective transition frequency remain below $1/(\tau\sqrt{N})$, implying tight requirements on $B$ field stability and cavity noise. First-order Zeeman couplings require sub-nT stability over the interrogation time, while second-order Zeeman couplings relax this to the $\mu$T range, as shown in optical clock experiments~\cite{2007NatPh...3..227B, Bloom:2013uoa, 2014arXiv1401.2378P, RevModPhys.87.637}. A configuration in which both Raman photons originate from the cavity field is parametrically weaker, since the effective coupling scales as $\Omega_{2\gamma}\propto E_{\rm cav}^2$ rather than linearly in the cavity field. We focus on the mixed Raman configuration as the more promising approach.

Figure~\ref{fig:reach} compares the projected reaches of these schemes with the sensitivities of existing and planned experiments, including bulk acoustic wave (BAW)~\cite{Goryachev:2014yra, Goryachev:2021zzn, Campbell:2023qbf, Campbell:2025mks} and cavity searches: FINUDA magnet for Light Axion SearcH (FLASH)~\cite{Alesini:2019nzq, Alesini:2023qed}, Axion Dark Matter eXperiment (ADMX)~\cite{ADMX:2018gho, ADMX:2019uok, ADMX:2021nhd, Chakrabarty:2023rha}, and Axion Longitudinal Plasma HAloscope (ALPHA)~\cite{Lawson:2019brd, ALPHA:2022rxj}. The light-shining-through-wall (LSW) Optical Search of QED vacuum magnetic birefringence, Axion and photon Regeneration (OSQAR-II)~\cite{OSQAR:2007oyv, Ballou:2014myz} and ALPS-II~\cite{Bahre:2013ywa, Beacham:2019nyx} are also shown, with bandwidth $\Delta f \approx f$ in Eq.~(124) of Ref.~\cite{Aggarwal:2025noe}. The differing slopes of the optical/NIR projections arise from changes in the effective bandwidth regime $\Delta f$ in Eq.~\eqref{eq:Deltaf_methods} and from the inclusion of squeezing in the aggressive configuration. These effects modify the frequency and atom number scaling of the strain sensitivity, and do not result from any assumed scanning strategy. The strain reach for optical/NIR readouts in Fig.~\ref{fig:reach} becomes numerically comparable to that of LSW experiments such as ALPS-II under the benchmark assumptions adopted here, although the underlying observables, detector responses, and search strategies are not identical. This result reflects the fact that, in the optical regime, sensitivity is mainly controlled by quantum optical readout rather than the mechanism generating the signal. In this sense, different detector architectures can approach comparable sensitivities even when based on distinct physical processes, as emphasized in recent analyses of graviton detection and quantum measurement~\cite{Carney:2023nzz}. For very weak EM fields, the minimum detectable power scales approximately as $P_{\rm min}\sim f/\tau$ in the broadband regime $\Delta f\sim f$, leading to a strain sensitivity $\sim (V/N/\tau)^{1/2}$. As a result, approaches based on graviton--photon conversion and LSW experiments can approach similar sensitivity levels despite their different underlying physics. These comparisons are indicative only: the atomic projections correspond to narrowband coherent sensitivities, while the laboratory bounds shown for comparison are derived under different bandwidth and signal assumptions.

Besides the schemes proposed, additional detection routes could be explored further. Highly charged ions such as He$^+$, Ne$^+$, or Ar$^+$, where inner-shell transitions lie in the extreme ultraviolet (EUV) at (10--100)\,eV, corresponding to HFGW frequencies $f\sim (10^{16}$--$10^{17})$\,Hz. For interrogation times $\tau=10\,\mu$s and EUV cavities with $Q\sim 10^2$--$10^4$, the resulting sensitivity is poor, $\sqrt{S_{h,{\rm min}}} \sim (10^{-10}$--$10^{-12}){\rm\,Hz^{-\frac{1}{2}}}$ in the conservative configuration. Although spontaneous scattering is negligible, these values remain far from astrophysical HFGW amplitudes, making such schemes primarily conceptual. Another approach is direct photon detection via photo-ionization: neutral atoms (Cs, Ar, Ne) or thin-film solids can be ionized once the converted photon energy exceeds threshold, e.g.\ $f\gtrsim 940$\,THz for Cs. Here no high-$Q$ cavity is required and decoherence plays no role, but the converted photon flux is extremely small. Even under optimistic assumptions for collection and readout efficiency, single-shot sensitivities are limited to $\sqrt{S_{h,{\rm min}}} \sim (10^{-20}$--$10^{-22}){\rm\,Hz^{-\frac{1}{2}}}$, useful mainly for constraining exotic scenarios rather than detection. Overall, microwave/Rydberg and coherent optical/NIR Raman schemes currently appear the most promising path, while EUV and photo-ionization approaches provide complementary coverage. Achieving astrophysical HFGW sensitivity in the EUV remains beyond present experimental capabilities.

\textbf{\textit{Discussion ---}} We first consider microwave/Rydberg implementations accessing the $(1$--$100)$\,GHz band. Superconducting cavities provide ultra-high quality factors and small mode volumes, while Rydberg atoms offer large dipole moments scaling as $\sim n^2$ and mature readout techniques via circuit QED and single-atom detection. Although the GW-induced signal is suppressed at lower frequencies within the present framework, the combination of robust cavity technology and strong atomic couplings makes the microwave approach a well-motivated platform for HFGW searches. The projected reach is comparable to other speculative concepts in this band, such as superconducting rings~\cite{Li:2009zzy} and enhanced magnetic-conversion setups~\cite{Anandan:1982is, Chiao:2002nv, Kiefer:2004hv}. Optical Raman schemes, such as those based on the Cs D2 line at $\lambda \approx 852$\,nm, operate at $f \sim (10^{14}$--$10^{15})$\,Hz. In this regime, the higher frequency enhances the GW-induced cavity response, while Raman and optical clock techniques enable long interrogation times $\tau \gg 1$\,ms with controlled spontaneous scattering for sufficiently large detuning $\Delta$. These features make optical implementations particularly promising under idealized conditions. The main experimental challenge is to combine high-$Q$ optical cavities with strong static magnetic fields and precise spatial overlap between the cavity mode and the atomic ensemble. In the most optimistic benchmark configurations, the simultaneous requirements of large magnetic fields, cavity volumes, coherence times, and quantum-limited atomic readout should be interpreted as technological targets rather than near-future detector designs.

The sensitivity estimates assume quantum-projection noise as the dominant limitation. In practice, technical noise overwhelms quantum noise many orders of magnitude earlier, setting the true experimental floor. In optical cavities, mirror-coating thermal noise, residual absorption, and mechanical vibrations broaden the linewidth and mask weak signals. In superconducting microwave cavities, magnetic-flux penetration and mechanical instabilities add dissipation and instability~\cite{Berlin:2023grv}. On the atomic side, laser phase and amplitude fluctuations, magnetic-field drifts, inhomogeneous broadening, atom-number uncertainty, and imperfect state preparation degrade coherence and contrast~\cite{Bringmann:2023gba, Kanno:2023whr}. Without suppressing such noise channels, shot-noise-limited sensitivities cannot be reached. Moreover, thermal and amplifier noise must be included in any realistic estimate~\cite{Berlin:2023grv}. In the microwave regime, blackbody photon occupation already dominates at $\sim 10$\,GHz unless the system cooled to sub-kelvin temperatures, while even the best amplifiers only approach the quantum limit under highly optimized conditions. Identifying and mitigating dominant noise mechanisms in each frequency band is crucial for prototype design. For EM conversion in a background field on resonance, the limiting strain scales with the EM energy $U$ stored in the cavity and the resonant $Q$ factor as $h_{\rm min} \propto U^{-1/2}Q^{-3/4}$~\cite{TitoDAgnolo:2024res}. The GW--EM transduction we propose is governed by the same physical setup, namely the magnetic field, geometry, bandwidth, and resonant enhancement. The additional improvement in Eqs.~\eqref{eq:Sh_Ryd_methods} and \eqref{eq:Sh_Raman_methods} comes from the atomic readout layer through collective shot-noise scaling with $N$ and $\tau$, not from the magnetic conversion stage alone, since the signal field is mapped into an atomic observable whose readout sensitivity improves with atom number, interrogation time and, possibly, additional squeezing. For this reason, the optimistic curves should be interpreted as cavity--atomic readout benchmarks rather than limits for direct EM power detection.

Our estimates also neglect cavity--atom hybridization~\cite{Kockum:2017sed, Kockum:2017xqr}. When the atomic transition lies close to the cavity resonance, the two systems form polariton modes with collective vacuum Rabi splitting $\Omega_N = 2g\sqrt{N}$, where $g$ is the single-photon atom--cavity coupling, modifying the signal transduction and noise properties. In this strong-coupling regime, the graviton-induced excitation populates a polariton mode rather than a bare cavity photon, changing the effective coupling strength and its scaling with atom number. The sensitivity estimates derived above assume instead a dispersive regime in which the atomic ensemble acts as a weakly perturbative readout rather than as part of the cavity eigenmode, requiring $g\sqrt{N} \ll 2\pi\,|f_{\rm at} - f|$, where $f_{\rm at}$ is the transition frequency between atomic levels and $f$ is the cavity resonance frequency. Strong hybridization occurs once the collective coupling exceeds the relevant dissipation scales,
\begin{equation}
    g\sqrt{N} \gtrsim 2\pi\,\max(\Delta f, \Delta f_{\rm at})\,,
\end{equation}
where $\Delta f_{\rm at}$ is the atomic linewidth. This issue is particularly relevant for Rydberg setups, where large transition dipoles and high-$Q$ cavities can naturally push the system toward the strong-coupling regime. For the optimistic benchmark parameters and $N\sim10^6$--$10^9$ atoms, one finds $g\sqrt{N}\sim10^5\,{\rm Hz}$. In practice, this can be addressed by operating the Rydberg ensemble as a dispersive electrometer rather than as a resonant absorber~\cite{Haroche:1989igc, Boss:2017ryq}, using MHz-scale detunings to remain in the perturbative regime. In optical Raman schemes, the larger optical linewidths make the dispersive interpretation less restrictive, even in the most optimistic configurations. A full coupled-mode treatment of the strongly hybridized regime is left for future work. Furthermore, only a fraction $\eta$ of the converted power couples into the resonant mode. Realistic $\eta<1$ depends on the magnetic field homogeneity, spatial overlap, cavity losses, and possible transfer losses between the conversion region and the atomic readout. Strong magnetic fields also cause Zeeman shifts $\propto \mu_B B$, where $\mu_B = 5.8\times 10^{-5},{\rm eV/T}$. For $B \gtrsim 1,{\rm T}$, these shifts can approach microwave transition frequencies, causing broadening and decoherence. A practical mitigation is to place the atomic ensemble in a magnetically shielded region, spatially separated from the conversion cavity and coupled via a waveguide, with the transfer loss absorbed into~$\eta$.

Overall, the proposed schemes provide strong projected sensitivity in two complementary HFGW bands under coherent narrowband signal assumptions. In the $(10^9$--$10^{11}),{\rm Hz}$ range, the microwave setup can approach benchmark coherent strain sensitivities in the most optimistic scenarios, potentially probing exotic early-Universe phenomena capable of generating sufficiently long-lived or spectrally narrow HFGW signals. Examples include resonant cosmic string emission~\cite{Servant:2023tua}, post-inflation relic backgrounds~\cite{Hindmarsh:2015qta, Hindmarsh:2017gnf, Guo:2020grp, Guo:2025cza}, and exotic compact object dynamics~\cite{Ireland:2023avg}. By contrast, chirping sources such as PBH mergers generally experience reduced resonant enhancement when their coherence time is shorter than the cavity response time entering $Q_{\rm eff}$~\cite{Dong:2015yjs, Carr:2020gox, Bavera:2021wmw, Zantedeschi:2024ram, Raidal:2024bmm}. The improved techniques may also probe inflationary and (p)reheating scenarios with characteristic amplitudes reaching $h^2\Omega_{\rm GW}\sim10^{-9}$~\cite{Bartolo:2015qvr,Graef:2015ova,Baumann:2007zm,Domenech:2021ztg,Cai:2020ovp,Cai:2023dls,Easther:2006gt,Garcia-Bellido:2007nns,Dufaux:2007pt}, although translating stochastic backgrounds into experimentally accessible narrowband strain sensitivities depends on coherence properties, bandwidth, and detector response.

\textbf{\textit{Conclusion ---}} We have presented a theoretical framework for detecting high-frequency gravitational waves (HFGWs) via graviton--photon conversion in strong magnetic fields, followed by readout using atomic quantum sensors. This work develops a cavity--atomic sensing framework for HFGW searches and shows how quantum sensing techniques could complement the broad range of detectors explored in the literature. We analyzed several concrete implementations, including Raman schemes in alkalis, microwave Rydberg transitions, photoionization thresholds, and inner-shell resonances, highlighting the versatility of atomic platforms across a wide frequency range. Depending on the configuration, the projected shot-noise-limited strain sensitivity spans from $\sqrt{S_{h, {\rm min}}} \sim 10^{-22}{\rm\,Hz^{-\frac{1}{2}}}$ in conservative microwave setups down to $\sqrt{S_{h, {\rm min}}} \sim 10^{-37}{\rm\,Hz^{-\frac{1}{2}}}$ in aggressive optical Raman benchmark scenarios. Under optimistic assumptions on cavity geometry, signal coherence, and quantum-limited readout, the projected narrowband coherent strain sensitivity can approach benchmark levels relevant for speculative HFGW scenarios, although realistic performance will depend sensitively on coherence, bandwidth, angular response, coupling efficiency, and technical noise.

These results are significant in two complementary ways. First, by approaching cosmological benchmark sensitivities in the GHz regime and reaching comparable levels in the optical band under optimistic assumptions, the schemes presented here illustrate the potential of HFGW searches to probe energy scales far beyond the reach of terrestrial colliders, opening a direct observational window onto high-energy physics. Second, the pursuit of HFGW detection drives advances in quantum sensing, motivating improvements in high-$Q$ cavity engineering, suppression of thermal and technical noise, control of large atomic ensembles, and long-lived coherence. Such advances have broad impact across precision measurement, quantum metrology, and fundamental tests of gravity. In summary, while astrophysical HFGW sources remain too faint to be detected, cascade cavity--atomic platforms provide a concrete framework for assessing whether cavity--atomic platforms could reach cosmological benchmarks and probe unprecedented frequency ranges.

\begin{acknowledgments}
We are grateful to Yifan Chen, Yuao Chen, Camilo Garc\'ia-Cely, Claudio Gatti, Yi Wang and Zhiwei Wang for insightful comments.
We also thank the Xplorer Symposia Organization Committee of the New Cornerstone Science Foundation for kindly inviting us to the symposium on Particle Physics on Tabletops (PPTT) in Chengdu, China, where this work has originated. 
This work was supported in part by the National Key R\&D Program of China (2021YFC2203100), by the National Natural Science Foundation of China (12350610240, 12433002, 12261131497), by CAS young interdisciplinary innovation team (JCTD-2022-20), by 111 Project (B23042), by CSC
Innovation Talent Funds, by the postdoctoral fellowship program of CPSF under grant number GZC20240212, by USTC Fellowship for International Cooperation, and by USTC Research Funds of the Double First-Class Initiative. 
LV acknowledges support by Istituto Nazionale di Fisica Nucleare (INFN) through the Commissione Scientifica Nazionale 4 (CSN4) Iniziativa Specifica ``Quantum Universe'' (QGSKY). LV further thanks the Tsung-Dao Lee Institute for its hospitality during the initial stages of this work.
\end{acknowledgments}

\bibliographystyle{apsrev4-1}
\bibliography{references.bib}

\end{document}